\documentclass{iopart}

\usepackage{graphics}
\usepackage{graphicx}
\usepackage{epsfig}

\begin{document}

\title{Real-space renormalisation group approach to driven diffusive
systems} \author{T. Hanney$^1$ and R. B. Stinchcombe$^2$}
\address{$^1$ SUPA and School of Physics, University of Edinburgh,
Mayfield Road, Edinburgh, EH9 3JZ, UK \\ $^2$Theoretical Physics, 1
Keble Road, Oxford, OX1 3NP, UK} 



\begin{abstract}
We introduce a real-space renormalisation group procedure for driven
diffusive systems which predicts both steady state and dynamic
properties. We apply the method to the boundary driven asymmetric
simple exclusion process and recover exact results for the steady
state phase diagram, as well as the crossovers in the relaxation
dynamics for each phase.
\end{abstract}

\section{Introduction}

Nonequilibrium phenomena are observed throughout nature and society,
and can often be thought of as some kind of driven diffusive system
\cite{SZ}. Examples include chemical kinetics, the motion of
interacting molecular motors along a substrate, population dynamics in
an ecology, the exchange of wealth or commodities in an economy,
traffic flow and many more. The ongoing interest in driven systems has
been sustained by the wide variety of behaviour the model systems
exhibit, which incorporate not only elements familiar from equilibrium
statistical mechanics, such as phase transitions, cooperative
behaviour, and so on, but also many dramatic new features. These include
phase transitions induced by external driving and even in one
dimension; long-range (i.e. power-law) correlations, not only at
criticality, but generically in the steady state; and very rich
dynamics.

Model driven diffusive systems are typically defined on a lattice on
which particles hop from site to site, where the precise definition of
the stochastic particle dynamics is motivated on physical
grounds. Nonequilibrium steady states are constructed by driving a
current of particles through the system. Despite being simply stated,
they still exhibit the wide range of phenomena. Indeed, a virtue of
their simplicity is that in some cases the models can be solved
exactly. However, most cases elude exact solution; moreover, even in
cases where the steady state can be written down, the task of
computing macroscopic observables may be too difficult. For these
reasons, it is necessary to develop approximate techniques. 

A technique that has been hugely successful in equilibrium statistical
mechanics is the renormalisation group procedure. Here, we show how one
can adapt these equilibrium methods to nonequilibrium systems, where
we devise a way to accommodate the feature that distinguishes the
nonequilibrium steady state from its equilibrium counterpart --- the
existence of currents --- in the renormalisation group
prescription. In particular, our approach includes the effects of
fluctuations, which, in certain equilibrium and nonequilibrium
systems, destroy the transitions predicted by mean field theory. Thus
our approach builds on previous prescriptions \cite{GM03}, and it
complements real-space techniques devised for quantum-spin chains, and
adapted to stochastic models \cite{HV, HS05}.

We test our approach using the asymmetric simple exclusion process
with open boundaries, which we introduce and review in section
\ref{ASEP}. In section \ref{MP}, we outline the matrix product
formulation of the master equation. This formulation provides a
convenient notation for the purposes of the steady state
renormalisation which we describe in section \ref{SSRG}. We recover
the exact critical point and find qualitative agreement for the phase
diagram. We extend the renormalisation to the dynamics in section
\ref{DRG} and again we obtain qualitative agreement with the exact
predictions for the relaxation dynamics.

\section{The asymmetric simple exclusion process (ASEP)} 
\label{ASEP}

The asymmetric simple exclusion process with open boundaries is one of
the most widely studied driven diffusive systems. Exact results show
that it undergoes both first- and second order phase transitions in
the steady state and exhibits different relaxation dynamics in each of
its phases. For these reasons, it represents an ideal test bed with
which to assess the efficacy of approximate methods.

The model is defined on a chain of $L$ sites, labelled
$l=1,\ldots,L$. The occupation variable $n_l = 0$ if site $l$ is
vacant and $n_l= 1$ if it is occupied; multiple occupancy is
forbidden. The stochastic dynamics are illustrated in figure \ref{ASEP
dynamics}: in continuous time, a particle at site $l<L$ hops to its
nearest neighbour site to the right with rate $p$, provided it is
empty. If site 1 is vacant, it becomes occupied with rate $\alpha$,
and if there is a particle at site $L$, it leaves the system with rate
$\beta$.
\begin{figure}
\begin{center}
\includegraphics[scale=0.4]{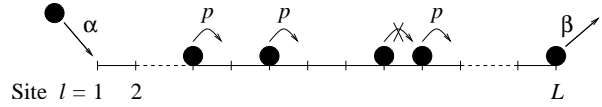}
\caption{Dynamics of the ASEP with particle injection at the left
  boundary and removal from the right boundary}
\label{ASEP dynamics}
\end{center}
\end{figure}
  
\begin{figure}
\begin{center}
\includegraphics[scale=0.4]{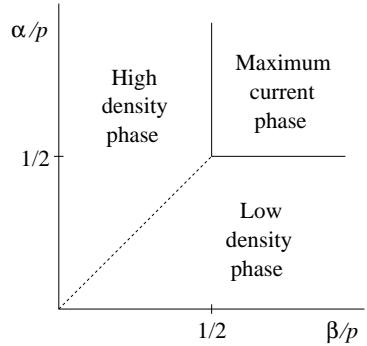}
\caption{Exact phase diagram for the boundary driven ASEP.}
\label{phase diagram}
\end{center}
\end{figure}

The phase diagram for this model is known exactly \cite{DEHP92}. In
the limit of large $L$, depending on the values of the ratios
$\alpha/p$ and $\beta/p$, the system may reside in one of three
possible phases (see figure \ref{phase diagram}): for $\alpha/p < 1/2$
and $\alpha < \beta$, the system is in a low density phase with
particle density $\rho < 1/2$; for $\beta/p < 1/2$ and $\alpha >
\beta$, the system is in a high density phase, $\rho>1/2$; for
$\alpha/p > 1/2$ and $\beta/p > 1/2$ the current, $J$,through the
system assumes its maximum value $J=1/4$ and the density $\rho = 1/2$.
The transition between the high and low density phases, indicated by
the dotted line in the phase diagram, is first order --- along the
transition line a low density region coexists with a high density
region and the system density $\rho = 1/2$. The transition between
these low current phases and the maximum current phase, indicated by
the solid line in the phase diagram, is second order.

The relaxation dynamics have also been obtained recently using the
Bethe ansatz \cite{deGE06}: in the maximum current phase the
relaxation time scales as $L^{z}$ where the dynamic exponent $z=3/2$;
along the coexistence line $z=2$; in the two low current phases away
from the coexistence line, the relaxation times are finite.

\section{Matrix product formulation of the ASEP}
\label{MP}
The exact solution for the steady state of the ASEP was obtained using
a matrix product technique \cite{DEHP92}. In this approach, the aim is
to represent the probability that one observes the system in a
configuration $\underline{n} = (n_1,\ldots,n_L)$ at time $t$, 
$P(\underline{n}, t)$, as
\begin{equation} \label{probabilities}
P( \underline{n}, t) = Z^{-1} \langle W | \prod_{l=1}^L ( n_l D +
  (1-n_l)E) |V\rangle\;,
\end{equation}
where $D$ and $E$ are (in general time-dependent) matrices, and
$\langle W |$ and $|V\rangle$ are (time-independent) vectors. Here,
$Z$ is a normalisation given by
\begin{equation} \label{normalisation}
Z = \langle W | C^L | V \rangle\;.
\end{equation}
where we have introduced the (time-independent) matrix $C=D+E$.  Thus,
a string of matrices is used to represent a configuration of the
particles in the system, where the matrix $D$ is used to represent an
occupied site and $E$ a vacant site.

\subsection{Steady state algebra}

In the limit $t\to \infty$, it was shown in \cite{DEHP92} that
(\ref{probabilities}) and (\ref{normalisation}) do indeed give the
steady state probabilities provided the matrices $D$ and $E$ and the
vectors $\langle W |$ and $|V\rangle$ satisfy the algebra
\begin{eqnarray}
D+E = DE\;, \label{bulk algebra}\\
\langle W|E = (p/\alpha)\langle W|\;, \label{lh boundary condition}\\
D |V\rangle = (p/\beta)|V\rangle \label{rh boundary condition}\;.
\end{eqnarray}
Moreover, matrix representations were found which satisfy (\ref{bulk
algebra})-(\ref{rh boundary condition}). We remark however that in
order to compute the probabilities to observe a particular
configuration, it is not necessary to find such matrix
representations; it is instead sufficient to use (\ref{bulk algebra})
to order all the matrices $E$ to the left of the matrices $D$, and
then use the boundary conditions (\ref{lh boundary condition}) and
(\ref{rh boundary condition}) to obtain probabilities in terms of
$\alpha$, $\beta$ and $p$.

\subsection{Dynamic algebra}

In \cite{SS95}, it was shown how the steady state algebra can be
generalised to a full dynamic algebra such that (\ref{probabilities})
and (\ref{normalisation}) represent the probabilities at all times
$t$. In this case, the matrices and vectors must satisfy
\begin{eqnarray}
\dot{D} = [ C^{-1}, \Lambda]\;,\label{eqn of motion}\\
\Lambda C^{-1}D =DC^{-1}\Lambda\;,\\
\langle W|(\alpha/p - C^{-1}\Lambda) = 0\;,\\
(\beta/p - \Lambda C^{-1}|V\rangle = 0\label{dynamic rh bc}\;,
\end{eqnarray}
where $\Lambda = DE$ is a current operator (capturing the fact that in
order for a particle to hop there must be a particle to the left of a
vacancy).

\section{Steady state renormalisation of the ASEP}
\label{SSRG}
The key feature which distinguishes nonequilibrium steady states from
equilibrium ones is the presence of currents. Moreover, in driven
diffusive systems with open boundaries, the boundary dynamics
determine the particle current and density throughout the bulk of the
system. This is in marked contrast to equilibrium systems where
boundary effects are typically localised to the vicinity of the
boundary. 

In this section, we present a renormalisation group method which
accounts for these currents by employing a blocking procedure designed
to accommodate the influence of the boundaries throughout the bulk of
the system.

\subsection{Blocking transformation}

\begin{figure}
\begin{center}
\includegraphics[scale=0.4]{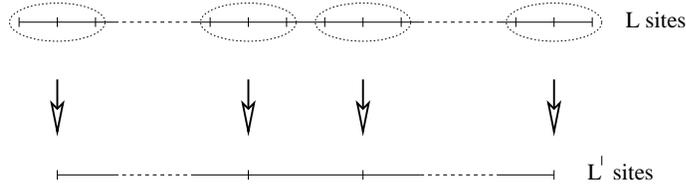}
\caption{Blocking the lattice into groups of $b=3$ sites.}
\label{blocking}
\end{center}
\end{figure}

A blocking transformation is implemented by grouping sites of the
lattice into blocks where each block contains $b$ sites, as
illustrated for the case $b=3$ in figure \ref{blocking}.
These blocks are then identified with the sites of a renormalised
lattice with a number of sites $L' = L/b$. This must be supplemented
by a prescription which identifies particle configurations within a
block with a single coarse-grained block variable. We achieve
this by using a majority rule identification, which for $b=3$ is
\begin{eqnarray}
D' = DDD + DDE + DED + EDD\;, \label{maj D}\\
E' = EEE + EED + EDE + DEE\;. \label{maj E}
\end{eqnarray}

\begin{figure}
\begin{center}
\includegraphics[scale=0.4]{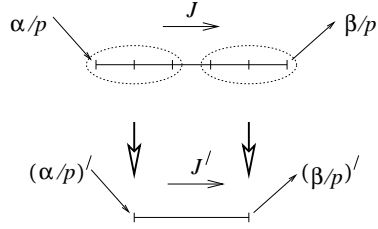}
\caption{Blocking transformation at fixed current for the boundary
driven ASEP with $b=3$.}
\label{asep blocking}
\end{center}
\end{figure}

The issue for any such renormalisation group treatment then is what
quantities are held fixed under the blocking and which are
rescaled. We keep the current across each bond $J$ and the bulk
density $\rho$ fixed, and rescale the parameters $\alpha/p$ and
$\beta/p$. To achieve this, we consider pairs of adjacent blocks
separately, and replace the effect of particle exchange with the rest
of the system with particle injection and ejection at the boundaries
of the pair of blocks, with rates $\alpha/p$ and $\beta/p$, as
illustrated in figure \ref{asep blocking}.
Thus the current and density in the rescaled system are written
\begin{eqnarray}
J' = \frac{\langle W' | D'E'|V'\rangle}{\langle W' |
  C'^2|V'\rangle}\;, \label{J'}\\
\rho' = \frac{\langle W' | D'|V'\rangle}{\langle W' |
  C'|V'\rangle}\;. \label{rho'}
\end{eqnarray}
Under the majority rule matching between configurations, (\ref{maj
  D}), (\ref{maj E}), the block current $J$ and the block density
  $\rho$ assume the form
\begin{eqnarray}
J = \frac{\langle W | (D^3 {+} D^2E {+} DED {+} ED^2 )( E^3 {+} E^2D {+}
  EDE {+} DE^2 ) | V \rangle}{\langle W | C^6 | V \rangle}\;,\label{J}\\ 
\rho = \frac{\langle W | (D^3 {+} D^2E {+} DED {+} ED^2) | V \rangle}{\langle
  W | C^3 | V \rangle}\;. \label{rho}
\end{eqnarray}
Note that we impose that neither the vectors and $\langle W |$ and
$|V\rangle$ nor the matrix algebra (\ref{bulk algebra}) to (\ref{rh
boundary condition}) are changed under the rescaling. This reflects
the fact that both the scaled and the unscaled configurations are
taken to evolve under the same stochastic dynamics: any extension of
the parameter space is prohibited under the rescaling. Finally, the
rescaling of the parameters $\alpha/p$ and $\beta/p$ is achieved by
setting $J' = J$ and $\rho ' = \rho$.

\subsection{$\alpha=\beta$}

Up to this point, the use of the matrix product formulation has been
nothing more than a choice of notation. The task now is to evaluate
the correlation functions (\ref{J'}) to (\ref{rho}) and this is
relatively straightforward if one makes use of the algebra (\ref{bulk
algebra}) to (\ref{rh boundary condition}), though we remark that even
here the algebra only simplifies the calculations: one could evaluate
the relevant correlations directly from the master equation, for
example. For the case $\alpha = \beta$ one obtains (still for $b=3$)
\begin{equation}
\rho' = \rho = 1/2\;,
\end{equation}
which is also the result obtained in the exact solution for $\alpha =
\beta$, and 
\begin{equation}
\frac{2x'}{3x'^2+2x'} =
\frac{66x+83x^2+46x^3+12x^4}{84x+126x^2+112x^3+70x^4+30x^5+7x^6}\;, 
\end{equation}
from matching $J' = J$, where $x = p/\alpha$. Stable fixed points of
this equation, $(\alpha/p)^* = (\alpha/p) = (\alpha/p)'$, are found at
$(\alpha/p)^* = 0$ and $(\alpha/p)^* = 3.959$, and these are separated
by an unstable fixed point at $(\alpha/p)^* = 1/2$. At the unstable
fixed point, $J^* = 1/4$. Thus the $(\alpha/p)^* = 0$ fixed point
represents the low-current phase, and the transition between this
phase and the maximum current phase is given by the unstable fixed
point, for which we recover the exact value $(\alpha/p)^* = 1/2$.
Moreover the current and density at this point are also given by their
exactly known values throughout the maximum current phase: $J^* = 1/4$
and $\rho^* = 1/2$.

\subsection{$\alpha \neq \beta$}

\begin{figure}
\begin{center}
\includegraphics[scale = 1.0]{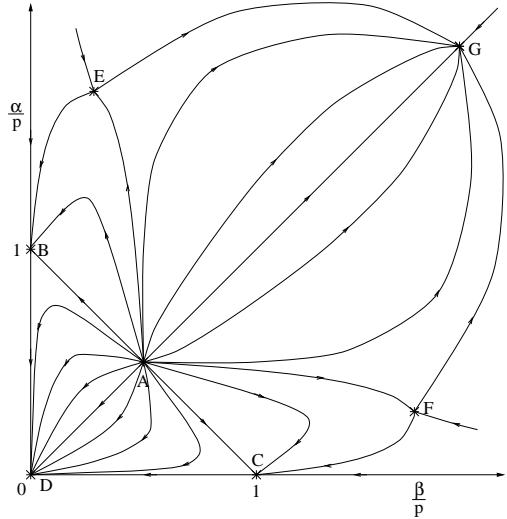}
\end{center} 
\caption{Schematic flow diagram in the $\alpha/p$-$\beta/p$ plane for
  generic values of $b$. The locations of the the fixed points A to D
  do not depend on $b$, the remaining fixed point values are
  $b$-dependent.}
\label{flow diagram}
\end{figure}

For $\alpha \neq \beta$ the scaling equations are straightforward to
obtain, but rather lengthy so we do not include them here. Their fixed
point structure is given in table \ref{fixed points}.
\begin{table}
\begin{center}
\begin{tabular}{|c|c|c|c|c|c|c|c|c|c|c|c|}\hline
 &A&B&C&D&E&F&G& & & &  \\
\hline\hline
$(\alpha/p)^*$ & $\frac{1}{2}$ & 1 & 0 & 0 & 3.248 & 0.423 & 3.959 &
0.415 & $\infty$ & 2.795 & $\infty$ \\
$(\beta/p)^*$ & $\frac{1}{2}$ & 0 & 1 & 0 & 0.423 & 3.248 & 3.959 &
$\infty$ & 0.415 & $\infty$ & 2.795 \\
\hline
\end{tabular}
\caption{Fixed points of the steady state blocking transformation for
$b=3$. The values given as decimals are evaluated numerically --- all
the rest are determined analytically. The fixed points depicted in the
flow diagram shown in figure \ref{flow diagram} are labelled A to G.}
\label{fixed points}
\end{center}
\end{table}
It is also straightforward to obtain the flow diagram by making some
choice for an initial set of parameters and seeing how they evolve
under several iterations of the scaling transformation. The result is
given in figure \ref{flow diagram}, and should be compared with the
exact phase diagram shown in figure \ref{phase diagram}.

The fully unstable fixed point A controls the second-order transition
between the low current and maximum current phases. At this point, the
current and density assume their exact critical values $(\rho_c, J_c)
= (\frac{1}{2}, \frac{1}{4})$. The flow line AE is a separatrix which
marks the boundary between the high density regime in the low current
phase (represented by point B) and the maximum current
phase. Similarly, the line AF separates the low density regime in the
low current phase (represented by point C) from the maximum current
phase. Points B, C and D are all zero current fixed points.  The
first-order transition in the model is represented by the flow line
AD, which separates (exactly) the high density from the low density
regimes in the low current phase. Also, across the line $\alpha/p +
\beta/p = 1$, the gradient of the density profile near the boundary
changes sign. This is the line BAC in the flow diagram.

Thus we have interpretations for the fixed points A to D. For the
remaining fixed points, we have no such physical interpretation,
though their presence is necessary for the flow diagram to adopt the
same qualitative structure as the phase diagram. Critical exponents
can be obtained by linearising the scaling equations in the vicinity
of the critical fixed point. Also, the scaling outlined above was
illustrated for blocking six sites to two ($b=3$). It is also possible
to perform the rescaling for eight sites to two ($b=4$) or nine sites
to three ($b=3$), for example, thereby increasing the block size or
the size of the renormalised clusters. Though these considerations do
not effect the topology of the phase diagram, they may improve
predictions for exponents. We illustrate this point more fully, for
dynamic scaling, in the next section.

\section{Dynamic scaling for the ASEP}
\label{DRG}
So far, the scaling has concerned only the statics. In this section,
we present a blocking transformation in order to learn about the
relaxation to the steady state. The blocking proceeds in the same
spirit as in the previous section and exploits the dynamic algebra
(\ref{eqn of motion}) to (\ref{dynamic rh bc}). Then the scaling
transformation, linearised in the vicinity of the physical steady
state fixed points (A to D in the flow diagram), informs us of the
nature of the relaxation dynamics in each of the corresponding phases.

\subsection{Blocking transformation}
We begin by replacing blocks containing two sites with a single
renormalised site (and will generalise subsequently). This single site
is parameterised by $(\alpha/p)'$ and $(\beta/p)'$ and its dynamics
are governed by the time evolution of $\langle D'(t) \rangle$:
\begin{equation} \label{one site}
\frac{\partial \langle D'(t) \rangle}{\partial t} = 
 \frac{\langle W' | \dot{D}'(t)|V'\rangle}{\langle W' | C'|V'\rangle}\;,
\end{equation}
(c.f. equation (\ref{rho'})). Again, the coarse-grained block variable
is identified with configurations in the original system through a
majority rule statement which, for blocks containing two sites,
assumes the form
\begin{equation} \label{majority rule}
D' = DD+\textstyle\frac{1}{2}(DE+ED) = \textstyle\frac{1}{2}(DC+CD)\;.
\end{equation}
This, carried into (\ref{one site}), yields the evolution in the
blocked system:
\begin{equation} \label{two sites}
\frac{\partial \langle D'(t) \rangle}{\partial t} = 
\frac{\langle W \vert \left[ \dot{D}(t) C + C\dot{D}(t) \right] \vert V
\rangle}{2 \langle W \vert C^2 \vert V \rangle}\;. 
\end{equation}
Now the dynamic algebra is used to compute the time evolution of the
relevant correlation functions. Although the rhs of (\ref{two sites})
depends only on $\langle DC \rangle$ and $\langle CD \rangle$, the
evolution of these two correlation functions depends on the third
time-dependent function $\langle DD \rangle$. Thus the time evolution
of the two-site block can be expressed as a matrix equation, where the
relaxation time is determined by the largest eigenvalue of a
$3\times3$ matrix. This eigenvalue gives the gap to the steady state
eigenvalue (which is zero), and we refer to this gap as
$\Delta^{(N_c)}$, where $N_c$ represents the number of sites in the
block: here, $N_c = 2$.

Now, under the coarse-graining, we hold the gap fixed therefore we set
\begin{equation} \label{matching}
\Delta^{(N_c')} = \Delta^{(N_c)}\;, 
\end{equation}
where $\Delta^{(N_c')}$ is the gap in the blocked system and $N_c'$ is
the number of sites in the coarse-grained system: here, $N_c' =
1$. Thus the dilation factor $b = N_c'/N_c$. Through equation
(\ref{matching}) we obtain a scaling equation in terms of the boundary
parameters. Because only a single quantity --- the gap --- is fixed
under scaling, we can derive a scaling relationship for a single
parameter only. Thus we consider the case $\alpha_p = \beta_p$.  This
is sufficient to probe the dynamics near the steady state fixed points
A and D.
The scaling relationship between
$(\alpha/p)'$ and $\alpha/p$ is not in general linear, therefore it is
possible that the dynamics at the zero-current fixed point and those
at the critical fixed point are characterised by different
values of the dynamic exponent $z$. Linearising the scaling equation
near either of these fixed points obtains
\begin{equation}
\alpha_p ' = \lambda \alpha_p ,
\end{equation}
which for each fixed point specifies the associated eigenvalue and
exponent $\lambda = b^{-z}$.

\subsection{$\alpha = \beta$}
We now illustrate this procedure for $N_c = 2$ and $N_c' = 1$ in more
detail, in the case $\alpha = \beta$. The dynamic algebra (\ref{eqn of
motion}) to (\ref{dynamic rh bc}) applied to equation (\ref{one site})
yields
\begin{equation}
\langle \dot{D}(t) \rangle = (\alpha/p) ' - 2(\alpha/p)'\langle D(t)\rangle\;,
\end{equation}
and the gap is easily seen to be $\Delta^{(1)} = 2(\alpha/p)'$
where the prime is used to indicate a parameter in the coarse-grained
system. For the two-site system, the dynamic algebra applied to
(\ref{two sites}) leads to the evolution of the correlation functions
which can be represented as
\begin{equation} 
\partial_t \left( 
\begin{array}{c} \langle DD \rangle \\ \langle DC \rangle \\ \langle
  CD \rangle \end{array} \right) = 
\left( \begin{array}{ccc} -2\alpha/p  & 0 & \alpha/p \\
1 & -(1+\alpha_/) & 0 \\ -1 & 1 & -\alpha/p \end{array} \right)
\left( \begin{array}{c} \langle DD \rangle \\ \langle DC \rangle \\
\langle CD \rangle \end{array} \right) ,
\end{equation}
from which the gap is found to be $\Delta^{(1)} = \alpha/p$. Thus we
obtain the scaling equation
\begin{equation}
(\alpha/p)' = \frac{\alpha}{2p}\;,
\end{equation}
therefore $\lambda = 1/2$ so $z=1$. We remark again that one could
just as easily perform this scaling working directly with the master
equation: the operator algebra simply provides a neat formulation for
our purposes.

In this case the same dynamic exponent is obtained for zero-current
and critical fixed points, but this is peculiar to blocking from two
sites to one. In other cases, one obtains for $(\alpha/p)'$ a
complicated nonlinear function of $\alpha/p$. This function is
linearised in the vicinity of the steady state fixed points A and D in
order to compute the dynamic exponent in the phases they
represent. The results of such a linearisation, for several values of
$N_c$ and $N_c'$ are shown in table \ref{dynamic exponents}.
\begin{table} 
\begin{center}
\begin{tabular}{|c|c|c|}\hline
$N_c \rightarrow N_c'$ & $z$ at $\alpha/p = \beta/p = 0$ & $z$ at
  $\alpha/p = \beta/p = 1/2$\\
\hline \hline
2 $\rightarrow$ 1 & 1 & 1 \\
\hline
3 $\rightarrow$ 1 & 1.118 & 1.072 \\
3 $\rightarrow$ 2 & 1.319 & 1.196 \\
\hline
4 $\rightarrow$ 1 & 1.194 & 1.111 \\
4 $\rightarrow$ 2 & 1.388 & 1.221 \\
4 $\rightarrow$ 3 & 1.486 & 1.257 \\
\hline
5 $\rightarrow$ 1 & 1.249 & 1.133 \\
5 $\rightarrow$ 2 & 1.437 & 1.234 \\
5 $\rightarrow$ 3 & 1.531 & 1.265 \\
5 $\rightarrow$ 4 & 1.589 & 1.274 \\
\hline
Exact & 2 & 3/2 \\
\hline
\end{tabular}
\caption{Estimates of the dynamic exponent near the zero current and
$J=1/4$ fixed points (points A and D in Figure \ref{flow diagram}).}
\label{dynamic exponents}
\end{center}
\end{table}
We also show the exact values which have only recently been obtained
\cite{deGE06}. Our results appear to capture the crossover from
$z=3/2$ dynamics in the maximum current phase (represented by point A)
to $z=2$ dynamics along the coexistence line (represented by point D).

\subsection{$\beta = 1-\alpha$}
One can also obtain scaling equations for the case $\beta = 1-\alpha$
in order to probe the dynamics at points B and C in the flow diagram,
points which represent the high and low density phases,
respectively. One finds that at both B and C, $(\alpha/p)' =
(\alpha/p)$ for all $N_c$ and $N_c'$.  This implies there is no system
size dependence in the leading contribution to the gap for large $L$,
which is consistent with the exact prediction of finite relaxation
times in the low current phases away from the coexistence line
\cite{deGE06}.

\section{Conclusion}
We have developed a renormalisation group procedure which we have
applied to the boundary driven asymmetric simple exclusion process. In
the steady state transformation, exact results for the critical point
are recovered as well as a flow diagram that shares the qualitative
structure of the exact phase diagram. The dynamic scaling also
captures the crossovers in the exactly computed relaxation dynamics in
each phase of the ASEP.

These scaling methods seem to be widely applicable: they can be
generalised to higher dimensions, multiple particle species and
multiple site occupancy. Since only small clusters give, through the
scaling, even the highly correlated properties, the scaling does not
depend on the availability of exact results, and may be particularly
useful in cases where the exact steady state can be written down but
any calculation of macroscopic properties remains difficult. They are
also direct and straightforward to interpret, without requiring any
translation from quantum spins or abstract fields as in
field-theoretic methods.

\ack T. H. would like to thank the SUPA
and the EPSRC for support under programme grant  GR/S10377/01.


\end{document}